\documentclass[aps,prb,twocolumn,showpacs,preprintnumbers,amsmath,amssymb,superscriptaddress]{revtex4}

\usepackage{graphicx}
\usepackage{dcolumn}
\usepackage{bm}
\usepackage{figsize}

\begin{document}


\title{Unconventional Reentrant Quantum Hall Effect\\in a HgTe/CdHgTe Double Quantum Well}

\author{M. V. Yakunin}
\email[]{yakunin@imp.uran.ru}
\affiliation{M. N. Miheev Institute of Metal Physics UB RAS, 620137 Ekaterinburg, Russia}
\affiliation{Institute of Natural Sciences, Ural Federal University, 620000 Ekaterinburg, Russia}
\author{S. S. Krishtopenko}
\affiliation{Institute for Physics of Microstructures RAS, GSP-105, 603950 Nizhni Novgorod, Russia}
\affiliation{Laboratoire Charles Coulomb, UMR 5221 Centre National de la Recherche Scientifique, University of Montpellier, F-34095 Montpellier, France}
\author{S. M. Podgornykh}
\affiliation{M. N. Miheev Institute of Metal Physics UB RAS, 620137 Ekaterinburg, Russia}
\affiliation{Institute of Natural Sciences, Ural Federal University, 620000 Ekaterinburg, Russia}
\author{M. R. Popov}
\affiliation{M. N. Miheev Institute of Metal Physics UB RAS, 620137 Ekaterinburg, Russia}
\author{V.~N.~Neverov}
\affiliation{M. N. Miheev Institute of Metal Physics UB RAS, 620137 Ekaterinburg, Russia}
\author{B. Jouault}
\affiliation{Laboratoire Charles Coulomb, UMR 5221 Centre National de la Recherche Scientifique, University of Montpellier, F-34095 Montpellier, France}
\author{W. Desrat}
\affiliation{Laboratoire Charles Coulomb, UMR 5221 Centre National de la Recherche Scientifique, University of Montpellier, F-34095 Montpellier, France}
\author{F. Teppe}
\affiliation{Laboratoire Charles Coulomb, UMR 5221 Centre National de la Recherche Scientifique, University of Montpellier, F-34095 Montpellier, France}
\author{S. A. Dvoretsky}
\affiliation{Institute of Semiconductor Physics SB RAS, 630090 Novosibirsk, Russia}
\author{N. N. Mikhailov}
\affiliation{Institute of Semiconductor Physics SB RAS, 630090 Novosibirsk, Russia}

\date{\today}

\begin{abstract}
We report on observation of an unconventional structure of the quantum Hall effect (QHE) in a $ p$-type HgTe/Cd$_x$Hg$_{1-x}$Te double quantum well (DQW) consisting of two HgTe layers of critical width. The observed QHE is a reentrant function of magnetic field between two $i=2$ states (plateaus at $\rho_{xy}=h/ie^2$) separated by an intermediate $i=1$ state, which looks like some anomalous peak on the extra-long $i=2$ plateau when weakly expressed. The anomalous peak apparently separates two different regimes: a traditional QHE at relatively weak fields for a small density of mobile holes $p_s$ and a high-field QH structure with a $2-1$ plateau--plateau transition corresponding to much larger $p_s$. We show that only a part of holes, residing in an additional light hole subband in the DQW, participate in QHE at weak fields while the rest of holes is excluded into the reservoir formed in the lateral maximum of the valence subband. All the holes come into play at high fields due to a peculiar behavior of the zero-mode levels.
\end{abstract}

\pacs{73.21.Fg, 73.43.-f, 73.43.Qt, 73.43.Nq}
\maketitle

\section{\label{sec:level1}Introduction}

Quantum Hall effect (QHE) at a fixed carrier density manifests most typically as a sequence of plateaus in the Hall magnetoresistance (MR) $\rho_{xy}(B)$ located in a stair-like fashion at its monotonically increasing values $\rho_{xy}(B)=h/ie^2$ for integer and fractional values of $i$ \cite{QHE}. This picture is disturbed by a nonmonotonic reentrant behavior of $\rho_{xy}(B)$  for the integer QHE (RIQHE) in some special cases: for two-subband conductivity due to a Landau level (LL) or subband mixing \cite{Lee}; in a traditional double quantum well (DQW) under tilted magnetic fields \cite{Gusev07} where certain QH states are repeatedly destroyed as a function of the parallel field component $B_\parallel$ due to an oscillating behavior of the tunneling gap; and around the $i=7/2$ and 5/2 fractional QH states where RIQH states $i=3\leftrightarrow4$ and $i=2\leftrightarrow3$ are observed at extra-low temperatures \cite{Eisenstein,Deng,Kleinbaum} due to repeated transitions between the collective quantum liquid and pinned quantum solid bubble or stripe states \cite{Koulakov,Goerbig}. In this Paper we demonstrate a distinct RIQHE appearing in a complicated energy spectrum of the HgTe DQW that may be basically explained without recourse to the collective nature of the electronic phases.

A uniqueness of the energy spectrum of the HgTe QW and its strong dependence on the well width \cite{KonigJPSJ} make it suitable to construct various kinds of a nontrivial energy structure in a system of two HgTe layers separated by a thin Cd$_x$Hg$_{1-x}$Te barrier, \textit{i.e.} in a HgTe/CdHgTe DQW \cite{Ya16,SSK}. Different applications were predicted for this structure \cite{Michetti12,Michetti13} as well as an opportunity to study fundamental phenomena in new conditions \cite{Budich}. Experimentally, several remarkable novel features were found in quantum magnetotransport of the HgTe DQW of relatively wide HgTe layers, like a reentrant sign-alternating QH states, a possibility to enlarge and regulate the band overlap and the enhanced zero filling factor state \cite{Ya16}. In the present study we found that probably the most unusual features of quantum magnetotransport are manifest in a $p$-type HgTe/CdHgTe DQW with the width of HgTe layers close to the critical value $d\approx d_c=6.3\div6.5$~nm, when a Dirac energy spectrum is formed in a single HgTe layer \cite{KonigJPSJ}. In this case, a RIQHE is revealed \cite{YaJETPL} that has indications of switching between two states with different densities of mobile holes. The only analog we know of such a kind of switching between different net carrier densities with magnetic field was reported for a multi-quantum-well structure made within a GaAs parabolic quantum well \cite{Gusev10}, where the effect was attributed to a spatial redistribution of electrons between different wells. In our case, we show that the observed `switching' is an inherent property of the collective DQW spectrum.

\section{\label{sec:Sample}Samples and measurements}

DQW structures were grown by molecular-beam epitaxy on a (013) orientated  GaAs substrate above a series of subsequent buffer ZnTe and CdTe layers and consisted of two HgTe layers with thickness $d=6.5$~nm separated by a 3-nm  Cd$_x$Hg$_{1-x}$Te barrier with $x=0.71$ sandwiched between the layers with the same $x$; without deliberate doping. The sample was shaped by photolithography to form a double Hall bar with soldered contacts at the contact pads penetrating through both the HgTe layers. The field-effect transistors were fabricated with parylene as an insulator and aluminum as a gate electrode. We measured the longitudinal and Hall MRs, $\rho_{xx}(B)$ and $\rho_{xy}(B)$, at temperatures down to 0.3 K in fields up to 13 T. The measurements were performed on a series of ungated and gated samples cut from the same wafer: the obtained results are identical. The ungated samples show $p$-type conductivity.

\section{\label{sec:Spec}Energy spectrum}

The energy spectrum as well as LL pictures were calculated in the single-electron 8-band $k\times p$-approach with inclusion of terms describing the influence of elastic strain by solving the self-consistent system of Schr\"{o}dinger and Poisson equations in the potential profile of the DQW \cite{SSK}.

\begin{figure}[b]
\includegraphics[width=\columnwidth]{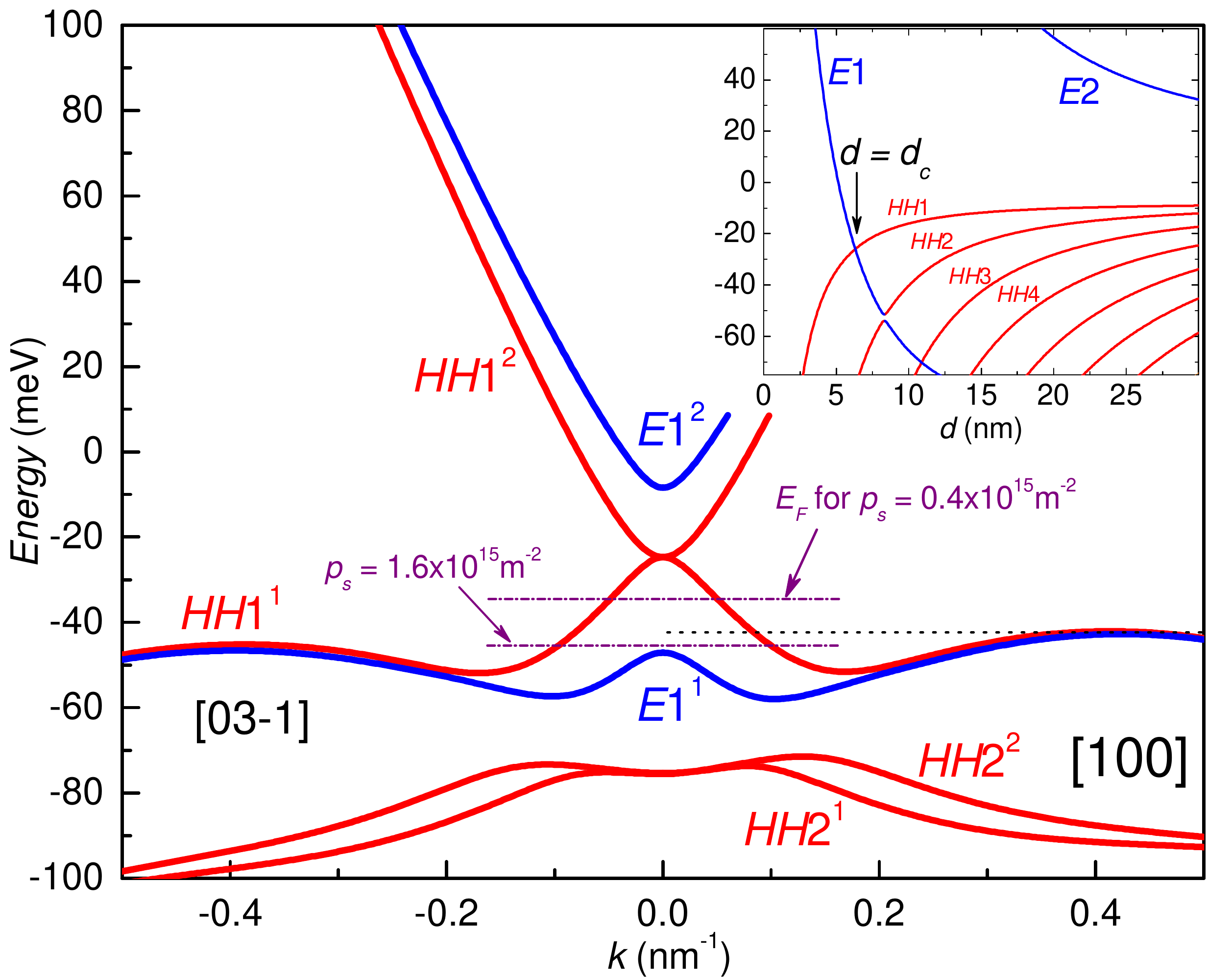}
\caption{\label{fig:E(k)} Energy spectrum of the symmetric DQW under study. Insert: spatial levels of a single QW as a function of the layer thickness $d$. The upper index on the main figure ($^1$ or $^2$) means the first or second split-off subband in a DQW originating from the corresponding level of a single well.}
\end{figure}

A picture of spatially quantized spectrum of a single HgTe QW is characterized by two groups of levels -- of the heavy hole $(HH)$ nature and light carriers $(E)$, see insert to Fig.~\ref{fig:E(k)}, -- that move quickly towards each other with increasing layer thickness $d$. At the critical thickness $d_c$ the extreme levels of both series cross and the Dirac spectrum is formed at this meeting point. It is clear from this picture that small deviations in $d$ will cause strong changes in the relative level positions. A DQW made of these HgTe has a specific energy spectrum as the $E$-levels are strongly tunnel-coupled, contrary to the $HH$-levels: Fig.~\ref{fig:E(k)}. Noteworthy is that the high sensitivity of the relative level positions to $d$ in a single QW means similar high sensitivity of the DQW spectrum to the sample structure.

\section{\label{sec:Exp}Experimental results}

\begin{figure}[b]
\includegraphics[width=\columnwidth]{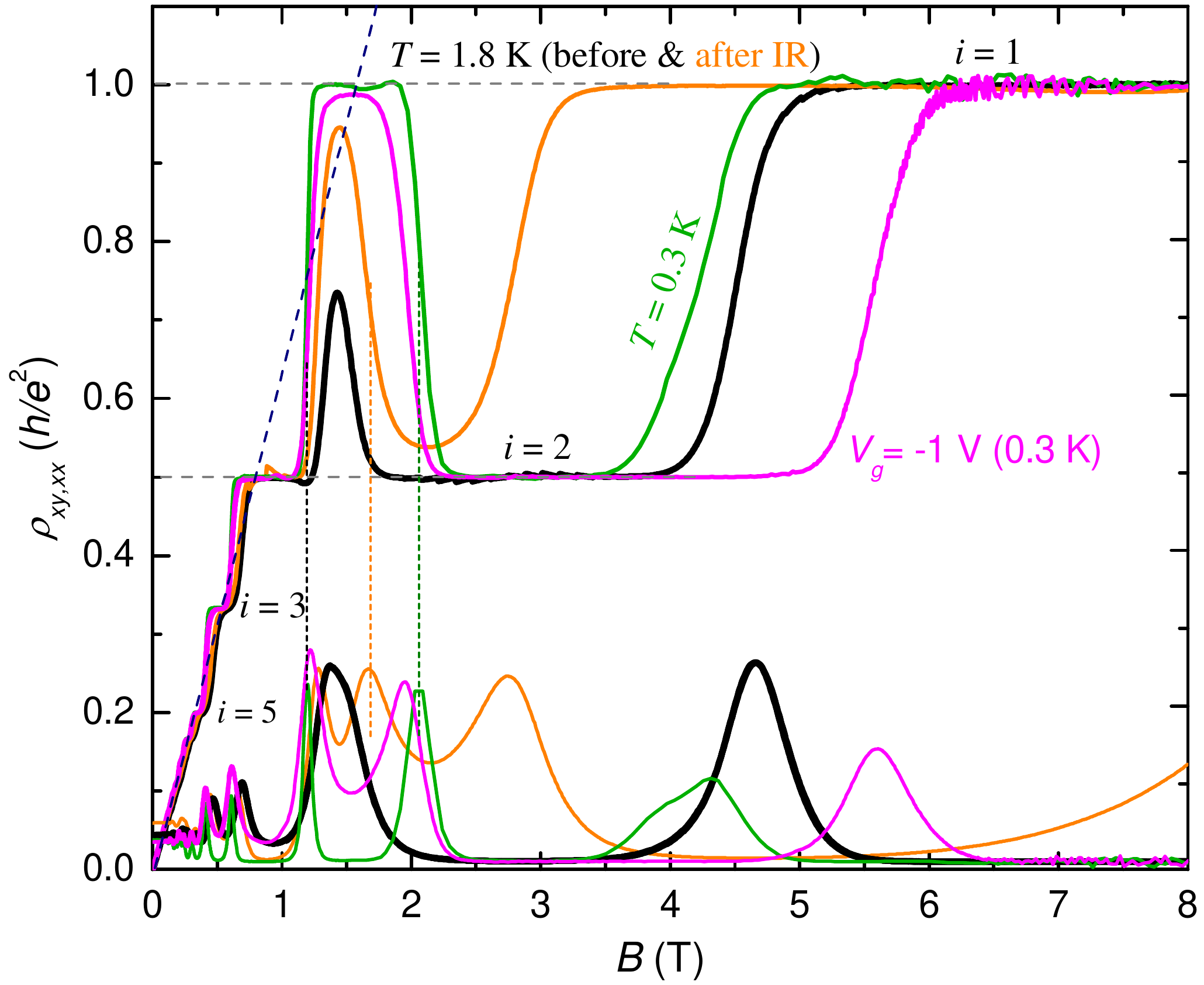}
\caption{\label{fig:rho(B)} MR with anomalous peak in $\rho_{xy}(B)$ before and after IR illumination, at 1.8 K and 0.3 K (unilluminated). Also the curve for $V_g=-1$~V at 0.3 K added.}
\end{figure}

\begin{figure}[t]
\includegraphics[width=\columnwidth]{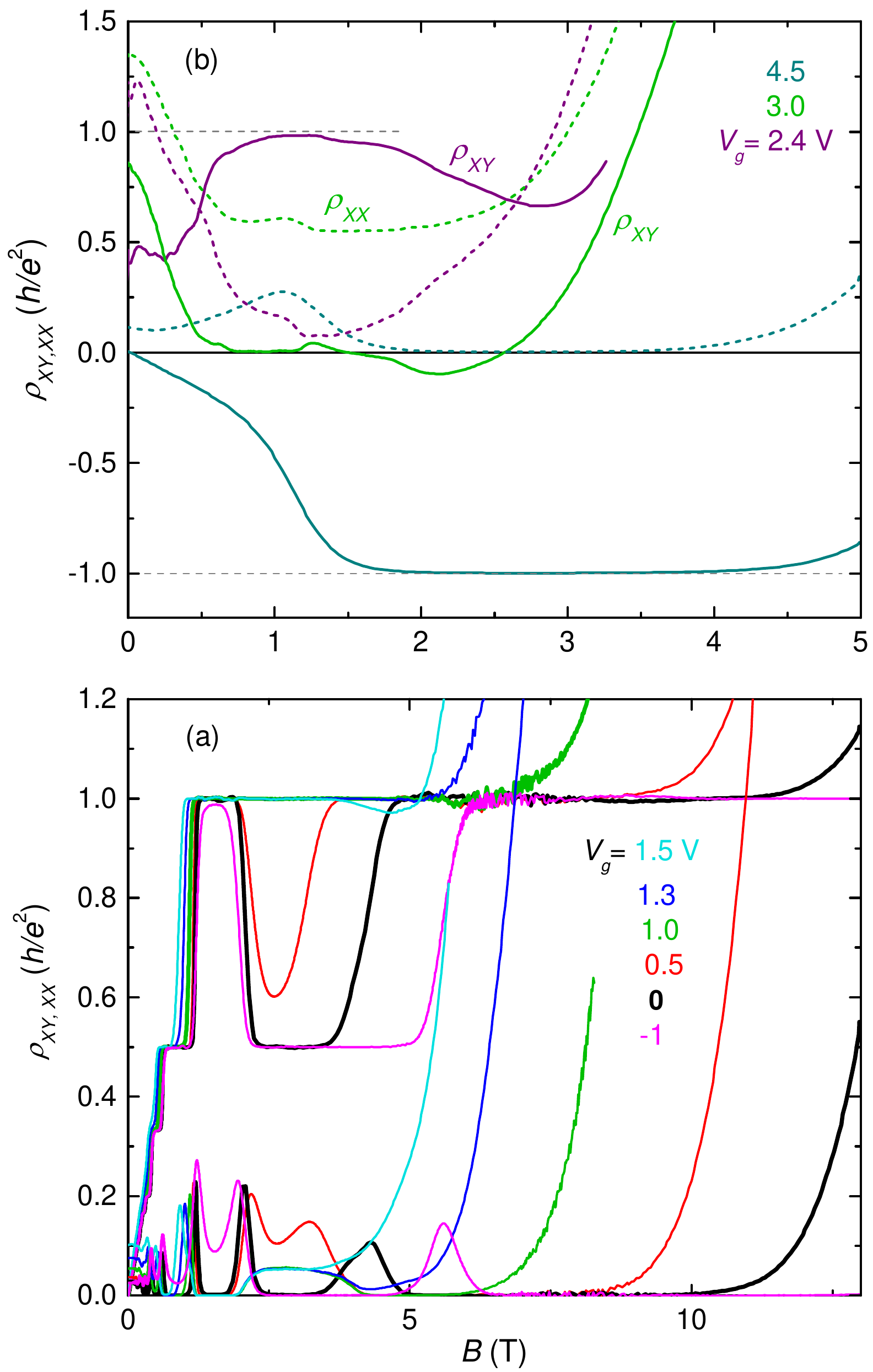}
\caption{\label{fig:rho(B,Vg)} Evolution of MR with the gate voltage: (a) $V_g=-1\div1.5$~V; (b) $V_g=2.4\div 4.5$~V. $T=0.3$~K.}
\end{figure}

We revealed a quite unusual structure of QHE in the described DQW (Fig.~\ref{fig:rho(B)}). In fields up to about 1.4 T, a traditional picture of QHE exists with plateaus of $\rho_{xy}(B)$ and transition regions between them centered around the straight line extrapolated from the classical Hall MR in weak fields. In this field region, $\rho_{xy}(B)$ corresponds to the two-dimensional hole density $p_s=0.4\times10^{15}$~m$^{-2}$. However, the further behavior of $\rho_{xy}(B)$ with increasing field is unusual: the incipient 2--1 plateau--plateau transition (PPT) reverts to the plateau $i=2$, which extends to an unexpectedly wide field interval. Thus, at temperatures above 1.8 K, a peak in the Hall MR of unknown nature is formed instead of the expected 2--1 PPT \cite{YaJETPL}. Further, the $i=2$ plateau does go over to the stable $i=1$ plateau, but the position of this 2--1 transition is shifted considerably to higher fields from the one extrapolated from the classical behavior of $\rho_{xy}(B)$ in weak fields. The hole density calculated for the field of this transition (\textit{i.e.}, for $i=1.5$) is $p_s=1.66\times 10^{15}$~m$^{-2}$. The estimated positions of the Fermi level, $E_F$, are presented in Fig.~\ref{fig:E(k)} for both values of $p_s$. Remarkable is that, for the latter value, $E_F$ is below the lateral maximum (LM) in the valence subband, thus, considering a high density of states (DOS) in LM, it should substantially influence the physics in this case.

The `anomalous' peak turns into a step at lower temperatures with its maximum turned into a short $i=1$ plateau. Thus the crux of this ‘anomalous’ peak is a transition into the $i=1$ state with the further reentrance back into the $i=2$ state.

The discovered anomalous peak in $\rho_{xy}(B)$, if it is narrow, is also seen as a similar peak in $\rho_{xx}(B)$ at the same field. But with development of the $\rho_{xy}(B)$ peak into the step with the $i=1$ plateau on its top, the corresponding peak in $\rho_{xx}(B)$ splits into two peaks positioned in the fields of the up and down slopes of this step. While the left peak of $\rho_{xx}(B)$ in this couple corresponds to a traditional 2--1 transition of $E_F$ through an extended state within a LL, the right one is something new and means in fact the crossing of the same (or similar) extended state in the backward direction. Also, $\rho_{xx}(B)$ exhibits a peak corresponding to the stable high-field 2--1 transition in $\rho_{xy}(B)$, which is positioned, as it should be, against its middle.
 
The weak- and high-field parts of MR behave quite differently in response to various impacts (Fig.~\ref{fig:rho(B)}): to IR illumination (that reduces $p_s$ persistently) and application of the gate voltage $V_g$. The high-field 2--1 transition reacts sharply manifesting the changes in $p_s$ while MR below the anomalous peak remains almost unchanged. Only for $V_g>+1$~V the whole MR curves start to change: (Fig.~\ref{fig:rho(B,Vg)}).

$\rho_{xy}(B)$  is distinctly antisymmetric with respect to the field polarity for $V_g<2$~V, whereas $\rho_{xx}(B)$ is symmetric, that excludes any relation between the nature of the observed anomalous peak and a mixing of two MR components as well as a presence of macroscopic inhomogeneities. This conclusion becomes invalid only around $V_g\approx 3$~V where a large growth of $\rho_{xx}(B)$ in the charge neutrality state creates problems in the unaveraged $\rho_{xy}(B)$ as is seen in Fig.~\ref{fig:rho(B,Vg)}(b).

\begin{figure}[t]
\includegraphics[width=\columnwidth]{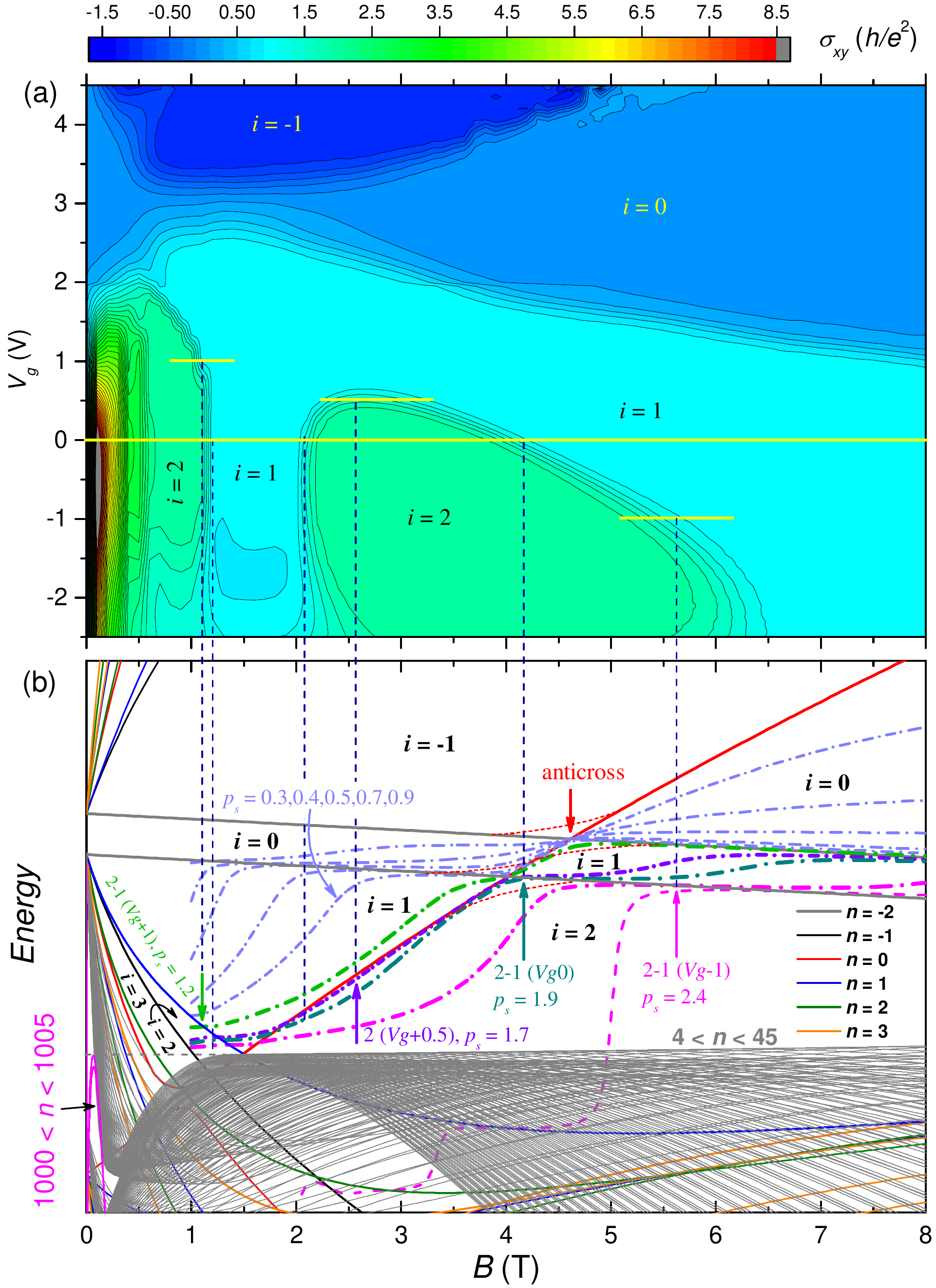}
\caption{\label{fig:S} (a) Evolution of the Hall conductivity $\sigma_{xy}(B,V_g)$ at $T=0.3$~K and its comparison with (b) the calculated picture of LLs. The latter is drawn for a fixed $V_g (=0)$ although it also changes with $V_g$. Dash-dot curves $E_F(B)$ are calculated for the given picture of LLs for a finite DOS and a set of net hole densities $p_s$ (indicated at the arrows in $10^{15}$~m$^{-2}$ together with the plateau numbers and $V_g$ in Volts). Short-dash red curves illustrate schematically supposed anticrossings between $n=0$ and $n=-2$ LLs. Arrows continued by dashed verticals connect characteristic features in the experimental map (indicated at certain $V_g$ by light-yellow horizontals), like $ i-j$ PPT or incipient $i$ plateau, with positions of $E_F(B)[V_g]$ \textit{vs.} corresponding LLs. The area segments are marked with $i$ values for the corresponding plateau numbers (negative for electrons). The dashed curve for $p_s=2.4$ is $E_F(B)$ for a hypothetical case of no LM.}
\end{figure}

The detailed evolution of QHE with $V_g$ is presented on Fig.~\ref{fig:S} for the Hall conductivity $\sigma_{xy}(B,V_g)$ (which is more convenient than MR since it does not go to infinity at the high fields). 

\section{\label{sec:Discus}Discussion}

To understand the observed singularities in QHE, the experimental data are compared in  Fig.~\ref{fig:S} to the calculated pattern of the DQW LLs with the plotted $E_F(B)$ curves. The latter were calculated for LLs of finite DOS and a set of fixed values of $p_s$, which in case of simultaneous population of the hole-like and electron-like LLs are the net values obtained as differences between the summary hole density and the electron density in these LLs \cite{Mendez,Suzuki}. It is important to distinguish correctly between the hole-like and electron-like LLs in such calculations. The issue is not trivial here due to anomalous behavior of certain LLs in the HgTe quasi-2D structures. An objective indicator of this attribution is a sign of the contribution from this LL to $\rho_{xy}$: it is positive for the hole-like and negative for the electron-like LLs. We determined it by the direction in which this LL shifts on approaching to the sample edge \cite{Hansen,Schmidt}: it is an electron-like (hole-like) LL if it shifts upward (downward). In all the calculations we are aware of (\textit{e.g.}, \cite{Schmidt,KonigSci}), this direction coincides with the direction of the LL shift with increasing magnetic field. Thus we use the latter criterion in our analysis regardless of what subband the LL comes from. Then, the two $n=-2$ zero-mode LLs \cite{Orlita,Schultz} in Fig.~\ref{fig:S}(b), slowly moving down with field, are the two topmost hole-like LLs and the $n=0$ LL, moving upward, is an electron-like LL, in spite of that it splits off from the valence subband.

For a symmetric DQW potential profile, the two $n=-2$ LLs should coincide, in agreement with the zero gap energy spectrum in Fig.~\ref{fig:E(k)}. The gap between two $HH1$ subbands (and between two $n=-2$ LLs) occurs with appearance of asymmetry in DQW. Existence of the $i=1$ QH plateau at the highest fields even for $V_g=0$ indicates that these two LLs are split and, hence, that the DQW potential profile is initially asymmetric \cite{SSK}. The asymmetry may be caused by imbedded charges on the surface or by geometrical deviations in DQW.

When an electron-like LL is imposed upon an array of the hole-like LLs, a special attention should be payed to the correct identification of filling factors in the gaps between LLs, to compare them with the corresponding QHE plateau numbers $i$. This numbering is trivial for the hole-like LLs below the electron-like LL as the latter is empty for $E_F$ positions in this part of the picture: Fig.~\ref{fig:S}(b). But a mixed electron-hole conductivity \cite{Mendez} appears for the energy range above the electron-like LL and below the topmost hole-like LLs. In this case the resulting filling factor (and the corresponding QH plateau number $i$) is a difference between the number of filled hole-like and electron-like LLs. This results in that the $i=1$ gap consists of two sectors in high and low fields separated by a neck, while the low field part of the gap between two $n=-2$ LLs corresponds to $i=0$. Similar situation occurs for the gaps with higher numbers $i=2, 3,\ldots$

The specificity of the mixed electron-hole conductivity was taken into account when calculating  $E_F(B)$ since not the pure hole density but the difference between the hole and electron densities is fixed (determined primarily by the balanced quantity of ionized impurities). When electrons appear in the electron-like LL then additional holes appear in the hole-like LLs in this balanced situation to preserve the electroneutrality.

The existence of lateral maxima in the valence subband in vicinity of the achievable $E_F$ positions creates quite a new situation here. A very large DOS in it is displayed in a dense grid of its LLs: Fig.~\ref{fig:S}(b). Noteworthy, to consider its influence in calculations of $E_F(B)$, a sufficiently big number of harmonics should be included ($n<45$ in Fig.~\ref{fig:S}(b)). The upper envelope of this grid is at the LM energy and extrapolates to it for $B\rightarrow 0$ and $n\rightarrow\infty$ (see LLs for $n=1000\div1005$ in Fig.~\ref{fig:S}(b)). This consideration clarify the issue that the LM LLs should start from the LM energy at $B=0$: it is an envelope of infinite array of LLs that obey this rule.

When, on decreasing the field, $E_F$ tends to jump into a higher number LL (see the dashed step-like curve in Fig.~\ref{fig:S}(b) for $E_F(B)$ in a hypothetical case of missing LM) it falls into the energy range next to LM with a high DOS within the tails of its LLs. As a result, $E_F(B)$ is stabilized within a narrow energy range close to LM. Important is that, being in the tail, $E_F$ does not cross the extended state of LM LL located at its center. On the contrary, the stabilized $E_F(B)$ crosses the LLs in the light hole fan-chart indicating each such crossing with a PPT in the experiment. But these crossings occur at much weaker fields than they would be in case of no LM (compare, \textit{e.g.}, 2--3 PPT for the dashed line at $\sim$3.5~T and  at $\sim$1~T for stabilized $E_F(B)$). Considering also that LLs in the LM grid are in fact unresolved due to their high density, it becomes clear that the LM LLs do not contribute to the structure of QHE in this range of fields despite that they absorb a considerable part of holes. Thus QHE is formed solely by the light hole LLs here.

This situation looks like a part of holes is localized in the states of LM at weak fields and only the other smaller part residing in the central maximum of the light hole branch (see Fig.~\ref{fig:E(k)}) is manifested in the experimental quantum features. With increasing field, $E_F(B)$ tends to reach the low number ($n=-2$) hole-like LLs, which are positioned much higher in energy than the rest of the hole-like LLs. Thus $E_F(B)$, on moving up in the wide $i=2$ gap, leaves the range of high DOS at LM, and now all the holes participate in formation of QHE. Hence the total hole density is manifested in the high-field 2--1 PPT. In this sense LM may be considered as a reservoir \cite{Zawadzki,Dorozhkin} absorbing a considerable amount of holes at weak fields and excluding them from QHE. Contrary to the previous works, this `‘reservoir' is not in a surrounding of the quasi-2D structure but within it being built into its own energy spectrum.

The superposition of the electron-like $n=0$ zero-mode LL upon a fan chart of light hole LLs (Fig.~\ref{fig:S}(b)) creates some novel configuration with a quasi-triangular gap of $i=1$  formed  below two topmost hole-like LLs and above the $n=0$ LL. This gap is surrounded by the gaps of $i=2$ to the left and to the right. Important is that the lower corner of this quasi-triangle is in the vicinity of the LM energy. In this case the stabilized $E_F$ cuts the lower corner of this triangle thus entering the $i=1$ gap in the limited range of fields and causing the step in $\rho_{xy}$ with $i=1$ plateau on its top superposed on the $i=2$ plateau. The corresponding behavior of $\rho_{xy}(B)$ for $V_g=0$ is highlighted in Fig.~\ref{fig:S}(a) by the long light-yellow horizontal and compared with the dash-dot $E_F(B)$ curve marked ($Vg0$) calculated for $p_s=1.9\times 10^{15}$~m$^{-2}$ in Fig.~\ref{fig:S}(b).

On decreasing $V_g$ from 0 to $-1$~V, the high-field 2--1 PPT moves to considerably higher fields (Figs.~\ref{fig:rho(B)}, \ref{fig:rho(B,Vg)}(a)) that is reflected in the general experimental picture of $\sigma_{xy}(B,V_g)$, Fig.~\ref{fig:S}(a), in a substantial slope of the border between the $i=2$ and $i=1$ phases at high fields. This corresponds to an increase of adjusted $p_s$ values (from 1.9 to 2.4)$\times10^{15}$~m$^{-2}$ in calculated curves of $E_F(B)$, Fig.~\ref{fig:S}(b). On the contrary, the forth and back transitions between the same QH phases in the weak-field range do not move with variation $V_g=0\rightarrow -1$~V and below in the experiment. Correspondingly, the low-field parts of the same $E_F(B)$ curves change little with $V_g$ as they get into the high DOS at LM. Only at the achieved extremely negative $V_g= -2\div -2.5$~V a narrowing of the $i=1$ step is seen in Fig.~\ref{fig:S}(a) indicating the approach of $E_F(B)$ curve to the lower corner of the $i=1$ quasi-triangle: Fig.~\ref{fig:S}(b). Some evolution of  the picture of LLs with $V_g$ is seen in Fig.~\ref{fig:S}(a) in that the width of the $i=1$ sloped stripe at high fields increases with negative $V_g$ manifesting an increase of the gap between the two $n=-2$ LLs.

On increase in $V_g$ from 0 to +0.5 V the $E_F(B)$ curve is gradually pushed out of the high-field $i=2$ gap (Fig.~\ref{fig:S}(b)) so that it only weakly enters it within a small range of fields for $V_g=+0.5$~V (see a short light-yellow horizontal at $2\div3$~T in Fig.~\ref{fig:S}(a)) resulted in a vanishing minimum on a quasi-continuous $i=1$ plateau (Fig.~\ref{fig:rho(B,Vg)}(a)). This minimum disappears at all for the higher $V_g$ as $E_F$ resides entirely in the $i=1$ gap. To be more exact, it resides in two sectors of this gap as is seen in Fig.~\ref{fig:S}(b). Surprisingly, passing of $E_F(B)$ through the neck between these two sectors does not manifest in $\rho_{xy}(B)$ although some bump appears in $\rho_{xx}(B)$ between $B=2$ and 4~T for $V_g=1\div 1.5$~V. These data mean that in fact there is an anticrossing between the zero-mode $n=-2$ and $n=0$ LLs so that a continuous connection is formed between these two sectors while the feature in $\rho_{xx}(B)$ indicates a local narrowing of the $i=1$ gap. Manifestations of anticrossings in QHE were found in a similar case of InAs/GaSb electron-hole system \cite{Zholudev} and in experiments on magneto-optics in the same HgTe/CdHgTe structures that we use. In the latter case the anticrossings were explained as being due to the bulk inversion asymmetry in this material \cite{Orlita,Zholudev,Bovkun} not considered in our calculations.

The lower-field border of the low-field 2--1 PPT still remains unchanged at positive voltages while $V_g<+1$~V (Fig.~\ref{fig:S}(a)), in agreement with a calculated weak raise of $E_F$ at $B\approx 1$~T with a $p_s$ decrease, and it is only with a further increase in $V_g$ that this PPT noticeably shifts to lower fields (Figs.~\ref{fig:rho(B,Vg)}(a),\ref{fig:S}(a)). Accordingly, the calculated $E_F$ shifts upward with a $p_s$ decrease below $1.2\times10^{15}$~m$^{-2}$ at $B\approx1$~T. It means that just at these small enough $p_s$ the Fermi level starts rising here, freed from being captured in the high DOS range of LM.

 A zero filling factor state $i=0$ is revealed as $\sigma_{xy}(B,V_g)\rightarrow0$ at high positive $V_g$ and high fields: Fig.~\ref{fig:S}(a). It means that the infinite growth of $\rho_{xy}(B)$ is delayed in fields with respect to the growth of  $\rho_{xx}(B)$: Fig.~\ref{fig:rho(B,Vg)}(a). A divergent shape of this $i=0$ state in the experimental $(B,V_g)$-map, Fig.~\ref{fig:S}(a), dramatically reproduces the shape of the high-field $i=0$ gap formed within the calculated picture of LLs, thus confirming our hierarchy of the gap numbering. The only problem is a quantitative disagreement between the calculated and experimental positions of the convergence point -- it is at $\sim$2~T in the experiment while at $\sim$4.5 T in the picture of LLs, Fig.~\ref{fig:S}(b). As reasons of this discrepancy may be: a similar anticrossing between LLs with the same numbers, the upper $n=-2$ LL and $n=0$ LL, that creates a tail for the upper $i=0$ gap extended to lower fields; an evolution of the picture of LLs with $V_g$ (while that in Fig.~\ref{fig:S}(b) is for a fixed $V_g=0$) and deviations of the DQW potential profile from the technologically preset one as well as some other unconsidered factors in calculations. A high sensitivity of the relative subband positions and, consequently, of the LL pattern to the smallest changes in the DQW potential profile has already been mentioned.
 
 With transition of $E_F(B)$ through the convergence point in decreasing field, it must pass from the $i=0$ divergent gap at high fields into a gap between two $n=-2$ LLs, which also corresponds to $i=0$ in our hierarchy. It is remarkable that the state $i=0$ is absent in the experimental map to the left from the convergence point within some field range. The functions $\sigma_{xy}(V_g)[B]$ taken at fixed fields manifest a pronounced $\sigma_{xy}=0$ plateau for the high fields but continuously pass from the $i=1$ to the electron phase with $i=-1$ without $i=0$ plateau at $B\approx0.5\div2$~T. The $\sigma_{xy}=0$ state reappears at the weakest fields but in fact the term `‘filling factor' is no more valid at this classical field range. The missing $i=0$ state at the fields below the convergence point means that the gap between two $n=-2$ LLs is absent at the corresponding $V_g=+3$~V and the DQW structure becomes symmetric. Thus, on application of $V_g$, our structure becomes more asymmetric with negative $V_g$ but symmetric for $V_g=+3$~V.

\begin{figure}[t]
\includegraphics[width=\columnwidth]{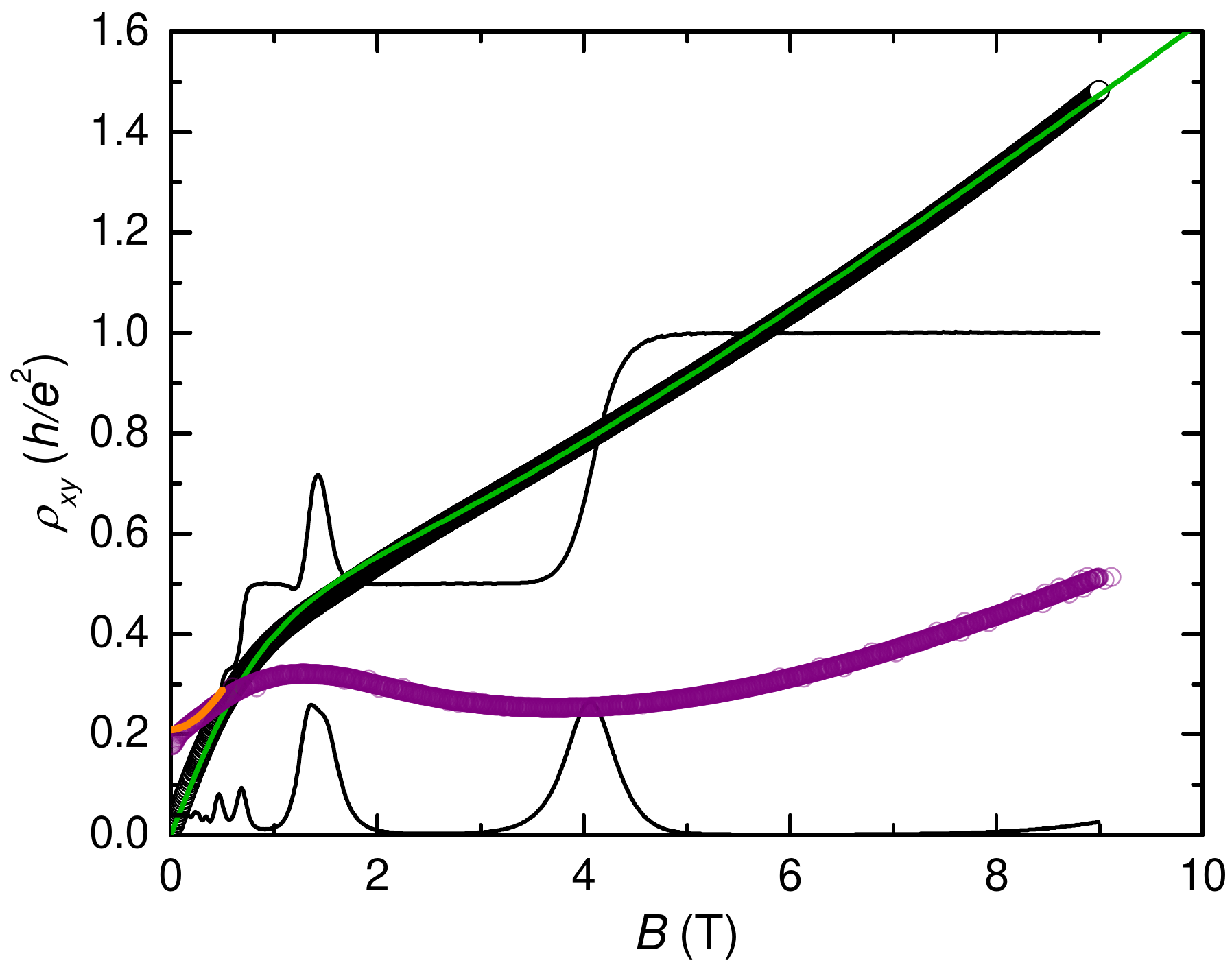}
\caption{\label{fig:QHE(T)} QHE at 1.8 K and MR at 30 K (circles) with a two-carrier fit for the latter (lines).}
\end{figure}

Observation of a distinct linear $\rho_{xy}(B)$ part as $B\rightarrow 0$ (Fig.~\ref{fig:rho(B)}) that yields $p_s$ much smaller than the value obtained at high fields may look controversial. To resolve this problem, QHE is compared with MR at 30~K when QHE is quenched: Fig.~\ref{fig:QHE(T)}. In these conditions, $\rho_{xy}(B)$ also consists of two parts with significantly different slopes, the low-field slope being the same as at low temperatures and the high-field part of $\rho_{xy}(B)$ crossing the high-field plateau for $i=1$. This high temperature $\rho_{xy}(B)$ is reproduced by classical magnetotransport of two kinds of holes, a  small amount of high-mobility holes: $p_1=0.27\times10^{15}$~m$^{-2}$, $\mu_1=3.6$~m$^2$/V$\cdot$s, and a large density of low mobility holes: $p_2=1.3\times10^{15}$ m$^{-2}$, $\mu_2=0.16$~m$^2$/V$\cdot$s. These densities are of the same order as those determined from QHE. As seen in Fig.~\ref{fig:E(k)} the two kinds of holes may originate from the central valence subband maximum and LM. 

\section{Conclusions}

Concluding, we can single out a combination of factors responsible for the formation of the observed anomalous QHE structure in the investigated DQW: (i) a lateral maximum in the valence subband on the background of a light-hole fan chart and proximity to it of $E_F$ in $p$-type structures; (ii) two topmost zero-mode hole LLs ($n=-2$) that are significantly raised in energy with respect to other hole LLs; (iii) a zero-mode electron-like LL ($n=0$) superimposed on the array of light-hole levels.

It is interesting to understand why the QHE peculiarities observed here are not found in a single  $p$-type HgTe layer with Dirac or inverted energy spectrum of a thickness close to critical (see, \textit{e.g.}, Ref.~\cite{Olbrich}, the LLs pictures for these cases may be seen in Ref.~\cite{Marcinkiewicz}). The three conditions listed above for observation of RIQHE seem to be realized here as well: LM not far from $E_F$, highly raised topmost hole LL and an electron-like $n=0$ LL. But the differences are that there is a single topmost hole $n=-2$ LL in a single QW, not two such levels as in DQW, and there is no additional array of light hole LLs that would create a background for the superimposed electron-like level. Due to the former, the corresponding filling factors and QH plateau numbers will be shifted one unit down in a single QW with respect to our case of DQW. As a result, $E_F(B)$ stabilization near LM would be developed in the gap with $i=1$, not $i=2$ as in a DQW. Then the approach of the Fermi level to the upper hole LL in a single QW will manifest in a different way since the $i=1$ QH plateau does not has a distinct high-field border (and an extra-long $i=1$ plateau was observed in some experiments \cite{Yahniuk}) while the high-field border for the $i=2$ plateau in the DQW is distinctly at $i=1.5$. Also, there is no cause for formation of an anomalous peak in a single QW since there is no additional light-hole array of LLs upon which the electron-like level could superimpose thus forming a quasi-triangular gap. The center part of the DQW energy spectrum resembles the fragments of the bilayer graphene spectrum around $K$ points \cite{McCann}, but the latter has no LM and no anomalous LLs \cite{McCann,Mireles}.

\begin{acknowledgments}
Authors are grateful to G.~M.~Min’kov and A.~A.~Sherstobitov for advice on making photolithography and for deposition of the gate. The research was carried out within the state assignment of the Russian Ministry of Education and Science (theme ``Electron'' No.~AAAA-A18-118020190098-5), supported in part by the Ural Branch of RAS Program (project No.~18-10-2-6) and RFBR (project No.~18-02-00172). Partially supported by MIPS department of Montpellier University through the "Occitanie Terahertz Platform", by the ANR project "Dirac3D", by the Occitanie region via the "Gepeto Terahertz platform", and by the CNRS through LIA "TeraMIR". A part of measurements was done in the Testing center for nanotechnology and advanced materials at M. N. Miheev Institute of Metal Physics UB RAS.\end{acknowledgments}

\end{document}